\documentclass[twocolumn,nopacs,prb,amsfonts,amsmath,amssymb,floatfix]{revtex4} 
\usepackage{color}
\usepackage{hhline}	
\usepackage{mathrsfs}
\usepackage{graphicx}
\usepackage{dcolumn}
\usepackage{bm}
\usepackage{multirow}
\usepackage{booktabs}
\usepackage{afterpage}
\usepackage{amsmath}
\usepackage{braket}
\usepackage{ulem}

\begin{document}

\title{First-principles study of defect formation energies in LaO$X$S$_2$ ($X=$ Sb, Bi)}

\author{Masayuki Ochi}
\author{Kazuhiko Kuroki}
\affiliation{Department of Physics, Osaka University, Machikaneyama-cho, Toyonaka, Osaka 560-0043, Japan}

\date{\today}
\begin{abstract}
We theoretically investigate defect formation energies in LaO$X$S$_2$ ($X=$Sb, Bi) using first-principles calculation.
We find that the oxygen vacancy is relatively stable, where its formation energy is higher in $X=$ Sb than in $X=$ Bi.
An interesting feature of $X=$ Sb is that the vacancy of the in-plane sulfur atom becomes more stable than in $X=$ Bi, caused by the formation of an Sb$_2$ dimer and the electron occupation of the impurity energy levels.
The formation energies of cation defects and anion-cation antisite defects are positive for the chemical equilibrium condition used in this study.
Fluorine likely replaces oxygen, and its defect formation energy is negative for both $X=$ Sb and Bi, while that for $X=$ Sb is much higher than $X=$ Bi.
Our study clarifies the stability of several point defects and suggests that the in-plane structural instability is enhanced in $X=$ Sb, which seems to affect a structural change caused by some in-plane point defects.
\end{abstract}

\maketitle

\section{Introduction}

High controllability is a key aspect of materials for systematic investigation of their properties.
From this viewpoint,
$Ln$O$XCh_2$ ($Ln =$ La, Nd, Ce, etc.; $X =$ Sb, Bi; $Ch=$ S, Se) is an important family of superconductors~\cite{BiS2,BiS2_2,Usui,BiS2review,BiS2review2,BiS2review3,BiS2review4,BiS2review5} and thermoelectric materials~\cite{LaOBiSSe_HP,BiS2_review_thermo,BiS2_review_thermo2,BiS2_review_thermo3}.
Their crystal structures consist of the conducting $XCh_2$ and insulating $Ln$O layers.
A rich variety of constituent elements and the existence of several siblings such as Bi$_4$O$_4$S$_3$~\cite{BiS2} and LaPbBiS$_3$O~\cite{LaPbBiS3O,LaPbBiS3O_2,LaPbBiS3O_F_thr,Kure}
are remarkable features of this family, which enable a systematic control of their electronic structure.
In $Ln$O$XCh_2$, electron carriers are usually introduced by fluorine substitutional doping for oxygen~\cite{BiS2_2}, or the valence fluctuation of cerium ions for $Ln=$ Ce compounds~\cite{Cevalence1,Cevalence2}.
Since atoms in insulating layers work as a charge reservoir in both ways, the charge carrier can be introduced without a large change in the electronic state of the conducting layers, which is also advantageous for studying their transport properties.

However, it was recently found that it is experimentally difficult to enhance the electrical conductivity in $X=$ Sb compounds against their robust insulating nature by the ways described above.
While several Sb compounds have been successfully synthesized such as Ce(O,F)SbSe$_2$~\cite{CeOFSbS2}, $Ln$OSbSe$_2$ ($Ln=$ La, Ce)~\cite{LnOSbSe2}, Ce(O,F)Sb(S,Se)$_2$~\cite{CeOFSbSSe2}, and NdO$_{0.8}$F$_{0.2}$Sb$_{1-x}$Bi$_x$Se$_2$~\cite{NdOFSbBiSe2}, 
it was also found that the electrical conductivity is low in these systems~\cite{CeOFSbS2,LnOSbSe2,CeOFSbSSe2}.
For example, it was reported that the electrical resistivity $\rho$ of CeOSbSe$_2$, CeO$_{0.9}$F$_{0.1}$SbSe$_2$, and LaO$_{0.9}$F$_{0.1}$SbSe$_2$ are $\sim10^0$--$10^1$ $\Omega$ m at room temperature~\cite{LnOSbSe2}. While it is smaller than that for LaOSbSe$_2$, $\rho \sim 10^3$ $\Omega$ m,  much lower resistivity is desirable for employing them as superconductors or thermoelectric materials.
Since our recent theoretical study predicts that high thermoelectric performance can be realized in $X=$ As, Sb compounds~\cite{BiS2Thermo_Dirac}, efficient control of the carrier concentration in Sb compounds is of great importance.
In fact, our previous theoretical calculation showed that the stability of the fluorine substitutional doping is lower in $X=$ Sb than in $X=$ Bi~\cite{Hirayama}.
However, because that study focuses on the relative stability between Sb and Bi compounds and assumes a simple fluorine-rich limit using a relatively small supercell, it is still unclear whether the fluorine substitutional doping is stable even in Sb compounds.
In addition, other possible point defects in mother compounds have not been theoretically investigated so far, even for Bi compounds, while they can affect the transport properties.
Because of the importance of the efficient carrier control in $Ln$O$XCh_2$ compounds, a systematic theoretical study on defect formation energies is highly awaited.

In this paper, we present a systematic investigation on point defects in LaO$X$S$_2$ ($X=$ Sb, Bi) by first-principles evaluation of their formation energies.
We find that anion replacements S$_\mathrm{O}$ and O$_\mathrm{S}$ are not stable, while V$_{\mathrm{O}}$ and V$_{\mathrm{S}}$ can take place.
The formation energy of V$_{\mathrm{O}}$ is higher in $X=$ Sb than in $X=$ Bi.
An interesting feature of $X=$ Sb is that the vacancy of in-plane sulfur becomes more stable than in $X=$ Bi by forming an Sb$_2$ dimer.
The formation energies of cation defects, $X_{\mathrm{S}}$, and S$_{X}$ are positive for the chemical equilibrium condition used in this study.
Fluorine likely replaces oxygen, and its defect formation energy is negative for both $X=$ Sb and Bi, while that for $X=$ Sb is much higher than $X=$ Bi.
Our study clarifies the stability of several point defects and suggests that the in-plane structural instability is enhanced in $X=$ Sb, which should be essential knowledge to understand and control the materials properties of LaO$X$S$_2$ and related compounds by impurity doping.

This paper is organized as follows. 
In Sec.~\ref{sec:cal}, we describe the calculation methods and computational conditions used in this study.
Calculated defect formation energies are shown in Sec.~\ref{sec:res}.
Section~\ref{sec:sum} is devoted to the conclusion of this study.

\section{Method\label{sec:cal}}

\subsection{Calculation of the defect formation energy\label{sec:calcdetails}}

By $N_1 \times N_2 \times N_3$ supercell calculation, the formation energy of a point defect $D$ in charge state $q$ is evaluated as~\cite{defect1, defect2}
\begin{align}
E_{\mathrm{form}}[ D^q ; \mathbf{N}] ( \Delta \epsilon_{\mathrm{F}})= E [ D^q ; \mathbf{N} ] + E_{\mathrm{corr}} [ D^q ; \mathbf{N} ]  \notag \\
- N_1N_2N_3 E_{\mathrm{P}}
 - \sum_i n_i \mu_i + q (\epsilon_{\mathrm{VBM}} + \Delta \epsilon_{\mathrm{F}} ),\label{eq:Ef}
\end{align}
where $\mathbf{N}=(N_1, N_2, N_3)$, $E [ D^q ;\mathbf{N}]$, $E_{\mathrm{corr}} [ D^q ;\mathbf{N}] $ and $E_{\mathrm{P}}$ represent the supercell size, the total energies of the supercell with the defect, its energy correction described below, and the total energy of the perfect unit cell without any defect, respectively. $n_i$ represents the number of removed (with a sign of $-$) or added (with a sign of $+$) atoms $i$, the chemical potential of which is denoted as $\mu_i$. 
$\epsilon_{\mathrm{VBM}}$ is the energy level of the valence-band maximum (VBM), and $\epsilon_{\mathrm{VBM}} + \Delta \epsilon_{\mathrm{F}}$ represents the Fermi level of the system.

In this study, we considered the energy correction $E_{\mathrm{corr}} [ D^q ; \mathbf{N} ] $ consisting of the following two terms:
\begin{equation}
E_{\mathrm{corr}} [ D^q ; \mathbf{N}]  = E_{\mathrm{pc}} (q;\mathbf{N})+ E_{\mathrm{be}}[ D^q ] .\label{eq:Ecorr}
\end{equation}
The first term $E_{\mathrm{pc}} (q;\mathbf{N})$ denotes the point-charge correction~\cite{Makov_Payne,defect1}, which was evaluated through the Ewald summation for screened Coulomb interaction in a periodic supercell consisting of a single point charge $q$ in a uniform background charge $-q$.
The screened Coulomb interaction is represented using the macroscopic static dielectric tensor in the way described in Ref.~[\onlinecite{diele_defect}].
As described later in this section, the remaining finite-size error is removed by the extrapolation with respect to the supercell size.

The second term $E_{\mathrm{be}}[ D^q ] $ in Eq.~(\ref{eq:Ecorr}) represents the band-edge correction, which is necessary when one uses a calculation method with a sizable band-gap error, such as the local density approximation and the generalized gradient approximation in density functional theory.
We applied the band-edge correction for a shallow defect level as described in Refs.~[\onlinecite{Zunger1,Zunger2}].
Suppose a point defect $D$ with a charge $q_0$ does not provide any carrier to the system (e.g., $q_0=1$ for $\mathrm{F}_{\mathrm{O}}$ that represents F$^{-}$ substitution for O$^{2-}$), and then we define
\begin{equation}
E_{\mathrm{be}}[ D^q ]  = \begin{cases}
(q-q_0) \Delta_{\mathrm{bg}}^{\mathrm{v}} \ \ (q> q_0)\\
(q-q_0) \Delta_{\mathrm{bg}}^{\mathrm{c}} \ \ (q\leq q_0)
\end{cases},
\end{equation}
where the band-edge correction to the valence band maximum $\Delta_{\mathrm{bg}}^{\mathrm{v}}$ and the conduction band minimum $\Delta_{\mathrm{bg}}^{\mathrm{c}}$ are evaluated by elaborate approximations in first-principles calculation that can provide an accurate band gap, as described in the next section.
A relative energy level from the band edge, where the formation energy curves of $D^q$ and $D^{q_0}$ intersect, is kept unchanged by this correction. 
However, this approximation is not necessarily valid when the impurity levels are not shallow.
Thus, we did not consider $E_{\mathrm{be}}[ D^q ]$ for that case.
In the following sections, we mention whether $E_{\mathrm{be}}[ D^q ]$ was applied for each defect.
In any case, the VBM energy in Eq.~(\ref{eq:Ef}) is defined including this correction:
 \begin{equation}
 \epsilon_{\mathrm{VBM}} = \epsilon_{\mathrm{VBM}}^0 + \Delta_{\mathrm{bg}}^{\mathrm{v}},\label{eq:VBMcor}
 \end{equation}
 where $\epsilon_{\mathrm{VBM}}^0$ is an uncorrected VBM energy. This modification of $\epsilon_{\mathrm{VBM}}$ changes the origin of $\Delta \epsilon_{\mathrm{F}}$ to the corrected VBM energy. We consider defect formation energies for $\Delta \epsilon_{\mathrm{F}}$ lying between the corrected band edges.

After calculating $E [ D^q ; \mathbf{N} ] + E_{\mathrm{corr}} [ D^q ; \mathbf{N} ] $ in Eq.~(\ref{eq:Ef}) for a set of supercells with several $\mathbf{N}$, we performed the least-squares fitting of $E [ D^q ; \mathbf{N} ] + E_{\mathrm{corr}} [ D^q ; \mathbf{N} ] $ by $c_0 + c_1N^{-1}$, where $c_0$ and $c_1$ are coefficients of this fitting, and $N=N_1N_2N_3$ (cf.~Ref.~[\onlinecite{defect1}] for the $N^{-1}$ (or equivalently, $\Omega^{-1}$ with $\Omega$ being the supercell volume) dependence). After determining $c_0$ and $c_1$, we took the $N \to \infty$ limit, i.e., simply took $c_0$, to get the thermodynamic limit.
By this procedure, we can get the defect formation energy at the thermodynamic limit,
\begin{equation}
E_{\mathrm{form}}[D^q]( \Delta \epsilon_{\mathrm{F}}) = \lim_{N_1,N_2,N_3\to \infty} E_{\mathrm{form}}[ D^q ; \mathbf{N}] ( \Delta \epsilon_{\mathrm{F}}),
\end{equation}
which shall be shown in the following sections.

\subsection{Computational conditions}

We used the Perdew--Burke--Ernzerhof parametrization of the generalized gradient approximation (PBE-GGA)~\cite{PBE} and the projector augmented wave (PAW) method~\cite{paw} as implemented in the {\it Vienna Ab initio Simulation Package}~\cite{vasp1,vasp2,vasp3,vasp4}. 
The core electrons in the PAW potentials were taken as follows: [He] for O and F, [Ne] for S, [Kr]$4d^{10}$ for La and Sb, and [Xe]$5d^{10}4f^{14}$ for Bi.
The spin-orbit coupling (SOC) was not included unless noted because a huge computational cost is required for supercell calculations including SOC.
We performed spin-unpolarized calculation unless noted, also due to high computational cost.
For ions with a closed-shell electronic configuration such as F$^{-}$, the spin polarization is expected to be unstable. Thus, our approximation likely works well for point defects involving such an ionic state. On the other hand, this is not always the case for some point defects, and we note that our approximation can cause some error in that case.
The plane-wave cutoff energy of 500 eV and the Gaussian smearing with the smearing width of 0.15 eV were used.
The structural optimization was performed until the Hellmann--Feynman force becomes less than 0.02 eV\AA$^{-1}$\ on each atom.
All the calculations were performed at zero temperature and zero pressure.

For calculating O$_2$ and F$_2$ molecules, we placed an isolated molecule in a $15\ \mathrm{\AA} \times 15\ \mathrm{\AA} \times 15\ \mathrm{\AA}$ cell, and only the atomic coordinates were optimized. Spin-triplet oxygen molecule was calculated using the spin-polarized calculation.
For calculating several solid compounds used for evaluating the chemical potential $\mu_i$ of each atom $i$, their crystal structures were optimized by the computational conditions described in the previous paragraph. A sufficiently fine $\bm{k}$-mesh was taken for each material. The lattice parameters and the atomic coordinates were optimized within a constraint of a fixed space group.

\begin{figure}
\begin{center}
\includegraphics[width=5.5 cm]{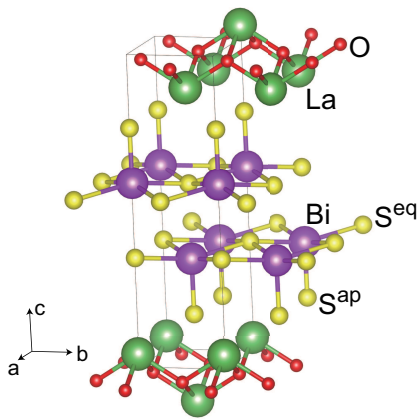}
\caption{The crystal structure of LaOBiS$_2$. Green, red, purple, and yellow spheres represent La, O, Bi, and S atoms, respectively. Lines shows the unit cell. All the crystal structures shown in this paper were depicted using the VESTA software~\cite{VESTA}.}
\label{fig:crystal}
\end{center}
\end{figure}

For supercell calculations of LaO$X$S$_2$ ($X=$ Sb, Bi), we first optimized both the atomic coordinates and the lattice parameters of the perfect crystal (i.e., without any defect), assuming the space group $P$2$_1$/$m$ (monoclinic).
This space group was realized in LaOBiS$_2$~\cite{SymLowPress,SymLow,BiS2_HP}. Figure~\ref{fig:crystal} shows the crystal structure of LaOBiS$_2$.
Also for Sb compounds, this space group was experimentally reported for Ce(O,F)SbS$_2$~\cite{CeOFSbS2} and Ce(O,F)Sb(S,Se)$_2$~\cite{CeOFSbSSe2}, and was theoretically suggested to be stable for LaO$X$S$_2$ ($X=$ As, Sb, Bi)~\cite{calc_P21m} and LaOSbSe$_2$~\cite{Hirayama}.
The space group is not necessarily common in $Ln$O$XCh_2$ compounds, because several BiS$_2$ compounds, such as CeOBiS$_2$~\cite{SymLow}, and also $Ln$OSbSe$_2$ ($Ln=$ La, Ce)~\cite{LnOSbSe2} were experimentally reported to have the tetragonal $P$4/$nmm$ space group.
Nevertheless, in this study, we can assume $P$2$_1$/$m$ in the calculation without loss of generality because this is a subgroup of $P$4/$nmm$, and found that the crystal structure indeed exhibits the monoclinic distortion as shown in Table~\ref{tab:lat}.

After calculating the perfect crystal, we performed the defect calculation by putting an isolated defect in the supercell with different sizes.
Here, we considered $3\times 3\times 2$, $4\times 4\times 2$, and $5\times 5\times 2$ supercells, which contain 180, 320, and 500 atoms, respectively.
For each supercell, we used $4\times 4\times 1$, $3\times 3\times 1$, and $3\times 3\times 1$ $\bm{k}$-meshes, respectively.
Supercell calculation optimized the atomic coordinates using the fixed lattice parameters optimized for the perfect crystal.
This treatment is justified because we are interested in the dilute limit of a defect.

For calculating the density of states (DOS), we used $24\times 24\times 4$ and $5\times 5\times 1$ $\bm{k}$-meshes for the primitive cell and the $4\times 4\times 2$ supercell, respectively.
We performed these DOS calculations using the fixed charge density obtained in the self-consistent-field (SCF) calculation.

\begin{table}
\caption{Optimized lattice parameters of LaO$X$S$_2$.}
\label{tab:lat}
\begin{tabular}{c c c c c}
\hline
 $X$ & $a$ (\AA)& $b$ (\AA) & $c$ (\AA) & $\beta$ ($^\circ$)\\
 \hline \hline
 \ Sb\ & \ 4.1018\ & \ 3.9949\  & \ 14.112 \ & \ 90.607\ \\
 \ Bi\ & \ 4.0725\ & \ 4.0505\  & \ 14.274 \ & \ 91.066\ \\
 \hline
 \end{tabular}
 \end{table}

We calculated the macroscopic static dielectric tensor based on the density functional perturbation theory to evaluate the point-charge correction. We considered both the electronic and the ionic contributions of the dielectric tensor. 
The local field effect for the Hartree and the exchange-correlation potentials was included. The derivative of the cell-periodic part of the Kohn--Sham orbitals were calculated using the finite-difference method using a $12\times 12\times 2$ $\bm{k}$-mesh.

For evaluating the band-edge correction, we performed band-structure calculation using the HSE06 hybrid functional~\cite{HSE06} including SOC.
Here, we adopt HSE06 because it reproduces the experimental band gap of LaOBiS$_2$ well: the direct band gap at the X ($=\mathbf{a}^* /2$) point calculated by HSE06$+$SOC is 0.97 eV, while the experimental optical gap is 1.0 eV~\cite{LaOBiS2_gap}.
Band-edge correction was evaluated as an energy difference $\Delta_{\mathrm{bg}} = \epsilon_{\mathrm{HSE06,\ SOC}} - \epsilon_{\mathrm{PBE,\ no\ SOC}}$ for the highest-occupied (for $\Delta_{\mathrm{bg}}^{\mathrm{v}}$) and lowest-unoccupied (for $\Delta_{\mathrm{bg}}^{\mathrm{c}}$) states at the X point, where $\epsilon_{\mathrm{HSE06,\ SOC}}$ and $\epsilon_{\mathrm{PBE,\ no\ SOC}}$ are the orbital energies obtained by HSE06 $+$ SOC and PBE without SOC calculations, respectively.
For the HSE06$+$SOC calculation, we used the crystal structures optimized by PBE calculation without SOC for a fair comparison of the orbital energies.
We used a $6\times 6\times 1$ $\bm{k}$-mesh for evaluating the band-edge correction.
Obtained band-edge corrections are shown in Table~\ref{tab:gapcor}.

While the band-edge correction was applied in the way described in the previous paragraph, the calculation error coming from PBE still remains in our calculation. For example, it is often said that GGA tends to delocalize the electronic state owing to the self-interaction error, which might result in an inaccuracy of the stability for the local structural change and a localized electronic state around it. Such an inaccuracy would be partially removed by using more sophisticate approximations such as the hybrid functionals for structural optimization, which is an important future issue.

\begin{table}
\caption{Band-edge correction (eV) for LaO$X$S$_2$.}
\label{tab:gapcor}
\begin{tabular}{c c c}
\hline
\ $X$\ & \ $\Delta_{\mathrm{bg}}^{\mathrm{v}}$\  & \ $\Delta_{\mathrm{bg}}^{\mathrm{c}}$\ \\
  \hline \hline
 \ Sb\ & \ $-0.24$\ & \ $-0.07$\ \\
 \ Bi\ & \ $-0.35$\ & \ $-0.23$\  \\
 \hline
 \end{tabular}
 \end{table}

\section{Results and Discussion\label{sec:res}}

\subsection{Chemical potentials of atoms\label{sec:chem}}

To evaluate the chemical potentials of atoms, $\mu_i$ in Eq.~(\ref{eq:Ef}), we determined possible sets of compounds that can coexist in chemical equilibrium in the following way.
Note that the chemical potentials of atoms are not uniquely determined because we can consider different experimental environments such as sulfur-rich and oxygen-poor ones.
In this section, we shall show all the possible equilibriums and then choose a representative one for which we shall present defect formation energies from the next section.

We first calculated the total energies of several solids consisting of La, O, F, Sb, Bi, and S atoms. 
Table~\ref{tab:list_mater} in Appendix~A shows a list of all the compounds considered here.
Next, we found out all possible subsets of the compounds, 
where every compound $\mathrm{La}_v \mathrm{O}_w \mathrm{F}_x X_y \mathrm{S}_z$ included in a subset satisfies
\begin{align}
E &[\mathrm{La}_v \mathrm{O}_w \mathrm{F}_x X_y \mathrm{S}_z] \nonumber\\
&= v \mu [\mathrm{La}] + w \mu [\mathrm{O}] +  x \mu [\mathrm{F}] + y \mu [X] + z \mu [\mathrm{S}],\label{eq:cond_atom}
\end{align}
while any compound not included in this subset satisfies
\begin{align}
E &[\mathrm{La}_v \mathrm{O}_w \mathrm{F}_x X_y \mathrm{S}_z] \nonumber\\
&>  v \mu [\mathrm{La}] + w \mu [\mathrm{O}] +  x \mu [\mathrm{F}] + y \mu [X] + z \mu [\mathrm{S}].
\end{align}
Here, the chemical potential of an atom is denoted as $\mu [\mathrm{A}]$ for atom A.
Each subset consists of five compounds, which enables consistent determination of the atomic chemical potentials using Eq.~(\ref{eq:cond_atom}).
Note that this procedure was done separately for $X=$ Sb and Bi.
These conditions represent the chemical equilibrium where five compounds included in a subset coexist while other compounds are unstable to exist.
For $X=$ Sb, we assumed that LaOSbS$_2$ is included in every subset.
On the other hand, for $X=$ Bi, we found that we cannot include LaOBiS$_2$ because LaOBiS$_2$ is slightly unstable with respect to the following reaction,
\begin{equation}
\mathrm{LaOBiS}_2 \to \frac{1}{2} \left( \mathrm{La}_2\mathrm{O}_2\mathrm{S} + \mathrm{Bi}_2\mathrm{S}_3 \right),
\end{equation}
by 20 meV. 
This slight instability might be a calculation error or perhaps means that LaOBiS$_2$ is metastable at zero temperature. 
It is difficult to discuss such a small energy difference and is outside the scope of this study.
Here, instead of including LaOBiS$_2$ into a subset, we just chose the chemical equilibriums where the instability of LaOBiS$_2$ is small, by discarding the subsets with $\Delta_{\mathrm{LaOBiS}_2}\equiv E [\mathrm{LaOBiS}_2] - (\mu [\mathrm{La}] + \mu [\mathrm{O}] + \mu [\mathrm{Bi}] + 2\mu [\mathrm{S}] ) > 100$ meV.
For both $X=$ Sb and Bi, we also imposed that every subset includes only one fluorine compound.
Thanks to this constraint, we can immediately obtain the chemical potentials of La, O, $X$, and S atoms satisfying the equilibrium condition without fluorine, by simply eliminating the fluorine compound from each subset.
This procedure yields the same chemical potentials of La, O, $X$, and S atoms regardless of the existence of fluorine, which is convenient to compare defect formation energies in chemical equilibrium with and without fluorine.

\begin{figure}
\begin{center}
\includegraphics[width=8.3 cm]{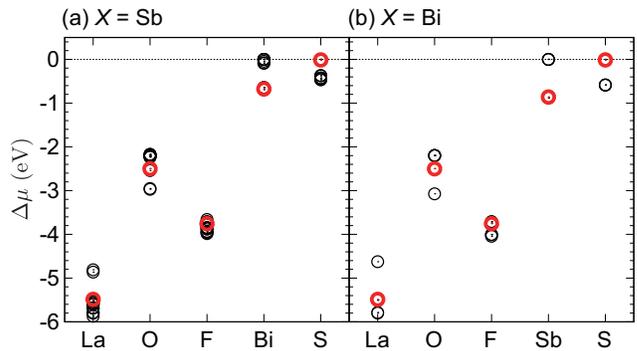}
\caption{Chemical potentials of atoms for (a) LaOSbS$_2$ ($X=$ Sb) and (b) LaOBiS$_2$ ($X=$ Bi). $\Delta \mu$ is defined in the main text. Red bold circles present the representative values that we used for showing defect formation energies. See the main text for details.}
\label{fig:chempots}
\end{center}
\end{figure}

All possible chemical potentials of atoms, to say, all possible chemical equilibriums, obtained here are listed on Tables~\ref{tab:equil_Sb} and \ref{tab:equil_Bi} in Appendix~A, for LaOSbS$_2$ and LaOBiS$_2$, respectively.
Figure~\ref{fig:chempots} summarizes these chemical potentials.
For this plot, we defined $\Delta \mu_i \equiv \mu_i - \mu_i^0$ for each atom $i$, where $\mu_i^0$ is the total energy per atom for La (solid), O$_2$ molecule, F$_2$ molecule, Sb (solid), Bi (solid), and $\alpha$-S (solid), for La, O, F, Sb, Bi, and S, respectively.
As a representative set, we chose the A1 (LaF$_3$, La$_2$S$_3$, La$_2$O$_2$S$_2$, La$_2$O$_2$S, and LaOSbS$_2$) on Table~\ref{tab:equil_Sb} for LaOSbS$_2$ and the B1 (LaF$_3$, La$_2$S$_3$, La$_2$O$_2$S$_2$, La$_2$O$_2$S, and Bi$_2$S$_3$) on Table~\ref{tab:equil_Bi}  for LaOBiS$_2$, the chemical potentials for which are shown by red bold circles in Fig.~\ref{fig:chempots}.
As shown in Fig.~\ref{fig:chempots}, these two chemical equilibriums provide similar values of chemical potentials for LaOSbS$_2$ and LaOBiS$_2$.
While we use these chemical potentials in the following subsections, we again note that the chemical potentials of atoms correspond to the experimental environment, which can be controlled by the experimental setup.
Thus, calculated defect formation energies presented hereafter can vary by differences in chemical potentials shown in Fig.~\ref{fig:chempots}.

Here we briefly mention the consistency between theoretically estimated chemical equilibriums and actual experimental environments for synthesis.
It was experimentally reported that La$_2$O$_2$S impurity is often found in the synthesis of LaOBiS$_2$ (e.g., Ref.~[\onlinecite{LaOBiS2_gap}]).
LaF$_3$ is used for introducing fluorine into LaOBiS$_2$ and is often found as an impurity phase (e.g., Ref.~[\onlinecite{BiS2_2}]).
Also, Bi$_2$OS$_2$, Bi$_2$S$_3$, and Bi impurity phases are found in the synthesis of Bi$_2$OS$_2$, which is a sibling compound of LaOBiS$_2$, where La is replaced with Bi in Bi$_2$OS$_2$~\cite{LaOBiS2_gap}.
Given these facts, it seems that our theoretical estimate to some extent represents experimental environments for synthesis.

\subsection{Anion point defects in mother compounds}

\begin{figure}
\begin{center}
\includegraphics[width=8.3 cm]{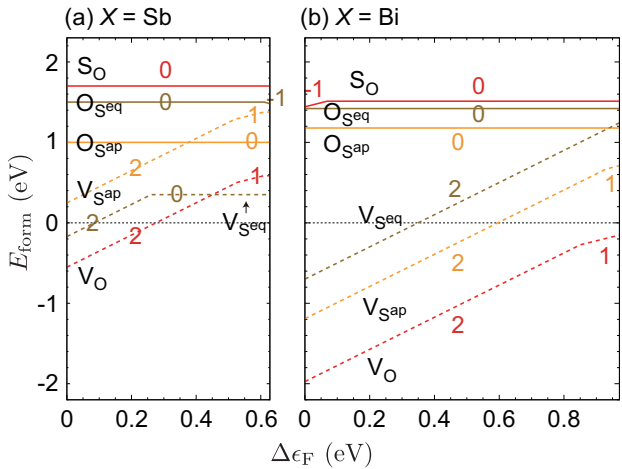}
\caption{Defect formation energies $E_{\mathrm{form}}$ of anion point defects for (a) LaOSbS$_2$ ($X=$ Sb) and (b) LaOBiS$_2$ ($X=$ Bi). The horizontal line (i.e. the Fermi level) is restricted to the energy range between the band edges corrected by HSE06. The values of $q$, which equals to the slope of each line, are shown beside the line.}
\label{fig:anion_both}
\end{center}
\end{figure}

\begin{figure*}
\begin{center}
\includegraphics[width=16cm]{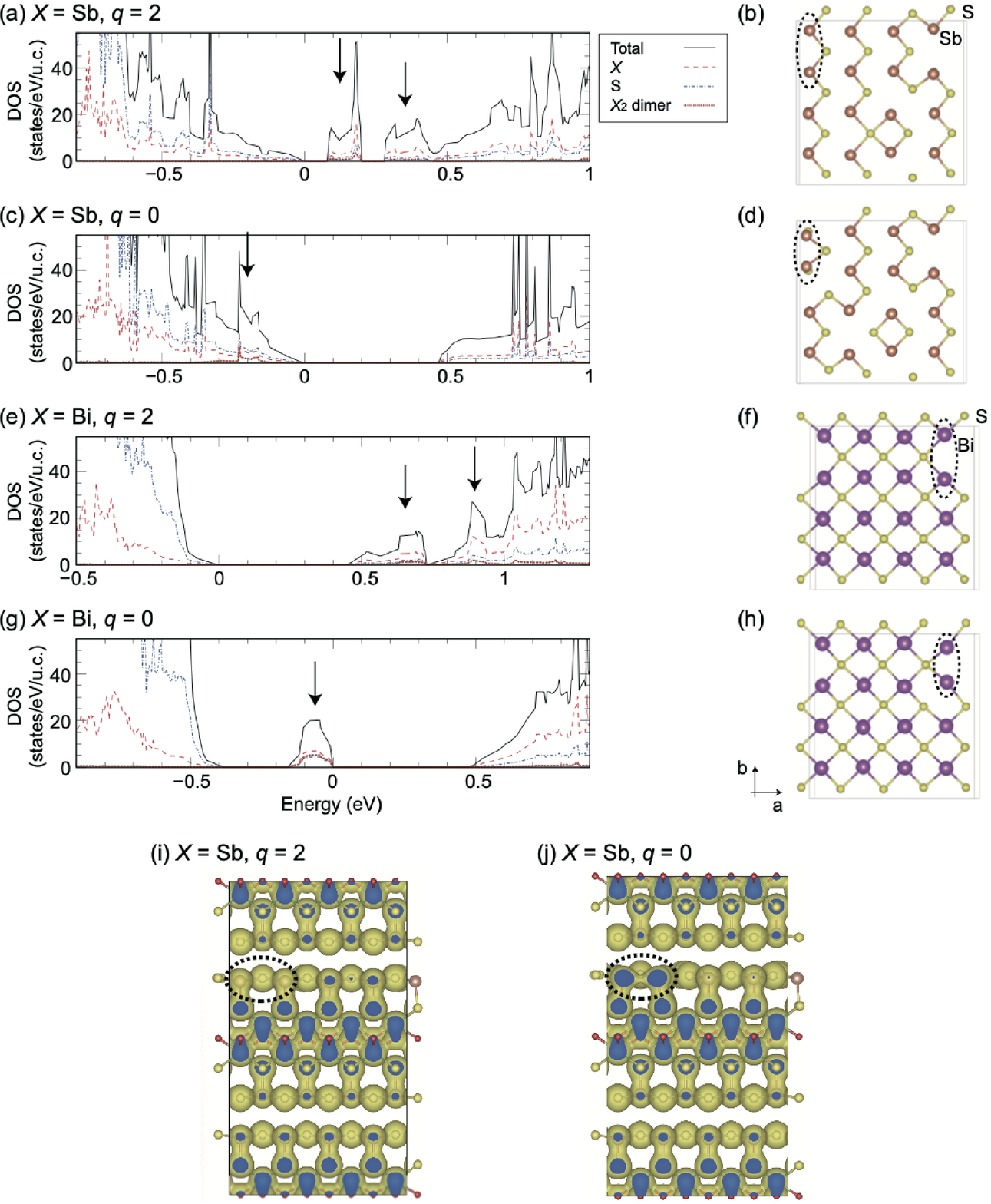}
\caption{(a)(c)(e)(g) Partial DOS and (b)(d)(f)(h) the crystal structure of the $X$S$_2$ plane that includes the $\mathrm{S}^{\mathrm{eq}}$ vacancy. (a)(b) $X=$ Sb with $q=2$, (c)(d) $X=$ Sb with $q=0$, (e)(f) $X=$ Bi with $q=2$, and (g)(h) $X=$ Bi with $q=0$. The zero of the energy in the DOS plots is the highest occupied energy level for each condition. Some of the impurity levels are shown with arrows in the DOS plots. The partial density of states for the $X_2$ dimer surrounded by the black broken lines in the crystal structure are shown in the DOS plots. In the crystal structure, brown, purple, and yellow spheres represent Sb, Bi, and S atoms, respectively. The valence electron density is shown for $X=$ Sb with (i) $q=2$ and (j) $q=0$. The black broken lines in panels (i)(j) correspond to those shown in panels (b)(d).}
\label{fig:Seqdef}
\end{center}
\end{figure*}

Formation energies of the anion point defects are shown in Fig.~\ref{fig:anion_both}.
In the figure, point defect species are denoted with the Kr{\"o}ger-Vink notation without showing the electronic charge since it varies against the chemical potential.  Lines with the most stable $q$ are shown for each chemical potential. For each line, a value of charge $q$ is shown in the figure.
Here, the slope of a line in the figure equals $q$ (see Eq.~(\ref{eq:Ef})).
The equatorial and apical sulfur atoms, S$^{\mathrm{eq}}$ and S$^{\mathrm{ap}}$, respectively, are defined in Fig.~\ref{fig:crystal}.
We applied the correction $E_{\mathrm{be}}[ D^q ]$ except V$_{\mathrm{S}^{\mathrm{eq}}}$, where the impurity energy level is not shallow, as we shall see later in this section.

In Fig.~\ref{fig:anion_both}, we can see that S$_{\mathrm{O}}$ and O$_{\mathrm{S}}$ are relatively unstable compared with V$_{\mathrm{O}}$ and V$_{\mathrm{S}}$.
For $X=$ Bi, the possible formation of O and S vacancies is consistent with the experimentally observed n-type carriers in the mother compound without fluorine doping. 
Note that here we adopted the S-rich environment, i.e., a high chemical potential $\mu [\mathrm{S}]$, as shown in Fig.~\ref{fig:chempots}, but V$_{\mathrm{S}}$ can be lowered by about 0.6 eV for $X=$ Bi when one adopts the S-poor environment as shown in Fig.~\ref{fig:anion_sppl}(c).
It is also noteworthy that the oxygen vacancy was experimentally observed in LaOBiSSe~\cite{LaOBiSSe_VO}.
For $X=$ Sb, V$_{\mathrm{O}}$ and V$_{\mathrm{S}^{\mathrm{ap}}}$ become much more unstable than $X=$ Bi, while V$_{\mathrm{S}^{\mathrm{eq}}}$ becomes much more stable at the conduction bottom with $q=0$ charge. We adopt the S-rich environment here, but the S-poor environment can lower the V$_{\mathrm{S}}$ formation energy by about 0.5 eV as shown in Fig.~\ref{fig:anion_sppl}(a), which makes the formation energy of V$_{\mathrm{S}^{\mathrm{eq}}}$ negative. Also, the formation energy of V$_{\mathrm{O}}$ can be almost zero by adopting the O-poor environment.

A negative formation energy for the whole energy range of the Fermi level within the band gap is troublesome from the viewpoint of the materials stability, because a large number of the point defects will be created in that case. We got such a low defect formation energy in some cases, such as V$_{\mathrm{O}}$ in Figs.~\ref{fig:anion_both}(b) and \ref{fig:anion_sppl}(c), V$_{\mathrm{S}}$ in Fig.~\ref{fig:anion_sppl}(a). One possible reason that makes this situation is the calculation errors, such as the finite-size error and the band-gap inaccuracy, both of which can in general result in an error in the formation energy of an order of 0.1 eV, even if one carefully pays attention to these issues and applies the correction terms. Another possibility is that some sets of the chemical potentials of atoms in Tables~\ref{tab:equil_Sb} and \ref{tab:equil_Bi} are not appropriate for materials synthesis. Considering the limitation of the accuracy of our calculation method, it is difficult to go into details further. Nevertheless, we again emphasize the consistency between our calculation and experiments where a small amount of electron carriers is observed.
If a defect that can introduce the electron carriers into the system has a negative formation energy for some Fermi level (e.g., near the valence band top), the Fermi level is automatically pushed up by the introduced electron carriers. Then, the Fermi level becomes close to the conduction band bottom, while the formation energy should be positive at the conduction band edge for the materials stability. Our calculation results well represent this trend.

Among the point defects investigated in this section, V$_{\mathrm{S}^{\mathrm{eq}}}$ is interesting in the sense that this defect offers an in-gap impurity state accompanied by a change of the (local) crystal structure.
Figure~\ref{fig:Seqdef} presents the partial DOS and the crystal structure of the $X$S$_2$ plane that includes the $\mathrm{S}^{\mathrm{eq}}$ vacancy.
We found that the crystal structures shown in Figs.~\ref{fig:Seqdef}(b)(f) are stable for $q= 2$, while those shown in Figs.~\ref{fig:Seqdef}(d)(h) are stable for $q=0,1$. The characteristic change here is the formation of the $X_2$ dimer in the $q=0,1$ structures as shown in Figs.~\ref{fig:Seqdef}(d)(h).
Here, the DOS peaks shown with arrows in Figs.~\ref{fig:Seqdef}(a)(c)(e)(g) have a relatively large component of the $X_2$ dimer. This tendency is clear for Fig.~\ref{fig:Seqdef}(g) while it is not clear for Fig.~\ref{fig:Seqdef}(e), because the Bi-Bi distance surrounded by the broken line in Fig.~\ref{fig:Seqdef}(f) (4.41 \AA) is much longer than that in Fig.~\ref{fig:Seqdef}(h) (3.42 \AA). A similar situation happens for the Sb compound, where the Sb-Sb distance is 4.08 \AA\ in Fig.~\ref{fig:Seqdef}(b) and 3.04 \AA\ in Fig.~\ref{fig:Seqdef}(d). The bond formation is usually caused by the electron occupation of the bonding orbitals that have low eigenenergies. In fact, the impurity levels with a relatively large $X_2$ dimer components denoted with arrows are occupied in Figs.~\ref{fig:Seqdef}(c)(g) ($q=0$), while it is not the case for the $q = 2$ states as shown in Figs.~\ref{fig:Seqdef}(a)(e). The valence electron density for $X=$ Sb with $q=0, 2$ shown in Figs.~\ref{fig:Seqdef}(i)(j) also supports the bond formation within such an $X_2$ dimer for the $q=0$ case.

It is also noteworthy that the $X$-S network is strongly disarranged in $X=$ Sb with $q=0$ as shown in Fig.~\ref{fig:Seqdef}(d).
This notable structural change might be related to the instability of in-plane atoms in BiS$_2$ compounds observed in experimental studies~\cite{BiS2_review_thermo2,BiS2_atomic_distortion}, which seems more significant in the case of $X=$ Sb.

\subsection{Cation point defects in mother compounds}

\begin{figure}
\begin{center}
\includegraphics[width=8.3 cm]{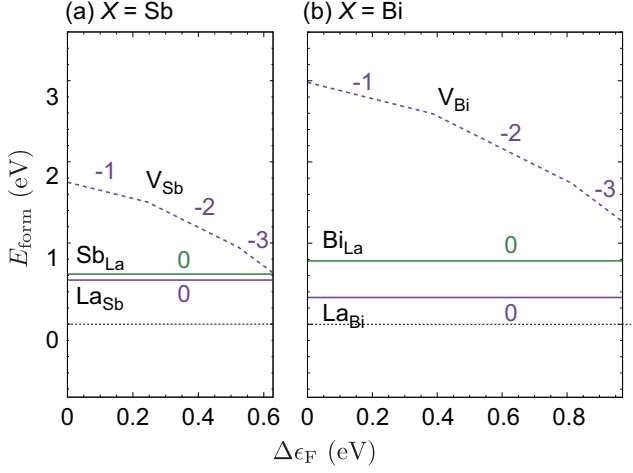}
\caption{Defect formation energies $E_{\mathrm{form}}$ of cation point defects for (a) LaOSbS$_2$ ($X=$ Sb) and (b) LaOBiS$_2$ ($X=$ Bi). The horizontal line (i.e. the Fermi level) is restricted to the energy range between the band edges corrected by HSE06. The values of $q$, which equals to the slope of each line, are shown beside the line.}
\label{fig:cation_both}
\end{center}
\end{figure}

Formation energies of the cation point defects are shown in Fig.~\ref{fig:cation_both}.
We do not show V$_{\mathrm{La}}$ here because we found that its formation energy is too high in the overall range of the chemical potential (e.g., $>4.5$ eV for $X=$ Bi).
We applied the correction $E_{\mathrm{be}}[ D^q ]$ except V$_{X}$, where the impurity energy level is not shallow, as we shall see later in this section.
All the defect formation energies shown in Fig.~\ref{fig:cation_both} are positive even at the conduction band bottom.

Here, we briefly discuss how the choice of the atomic chemical potentials can change our conclusion.
In Fig.~\ref{fig:cation_both}, we used an $X$-poor environment (see Fig.~\ref{fig:chempots} and Tables~\ref{tab:equil_Sb} and \ref{tab:equil_Bi}).
If one uses an $X$-rich environment, $\mu [\mathrm{Sb}]$ and $\mu [\mathrm{Bi}]$ can be increased by 0.68 and 0.86 eV, respectively, which destabilizes V$_{X}$ and La$_{X}$ and stabilizes $X_{\mathrm{La}}$. 
The energy diagrams for the $X$-rich conditions are shown in Figs.~\ref{fig:cation_sppl}(a)--(d).
As for the La chemical potential, we can adopt an La-rich environment but it also leads to a high $\mu [X]$ ($X$-rich, see A3 and A5 in Table~\ref{tab:equil_Sb} and B2 in Table~\ref{tab:equil_Bi}), which almost does not change the defect formation energies of $X_{\mathrm{La}}$ and La$_{X}$, as shown in Figs.~\ref{fig:cation_sppl}(a)(c).
In total, $X_{\mathrm{La}}$ can exhibit slightly negative formation energy only for an La-poorer and $X$-richer environment, as shown in Figs.~\ref{fig:cation_sppl}(b)(d).
Such a negative formation energy for the Fermi level in the whole energy region of the band gap is troublesome for the materials stability as discussed in the previous section. This situation is possibly due to the calculation error, such as the finite-size error and the band-gap inaccuracy, or means that that chemical environment is not appropriate for materials synthesis.
However, even when the system has a small number of $X_{\mathrm{La}}$ defects, because of the following three reasons, we do not consider that transport properties are much affected by this defect:
$X_{\mathrm{La}}$ is a defect in the insulating LaO layer rather than the conducting $X$S$_2$ layer, this defect introduces no carriers, and we do not find any in-gap impurity state for this defect.

\begin{figure}
\begin{center}
\includegraphics[width=8.3 cm]{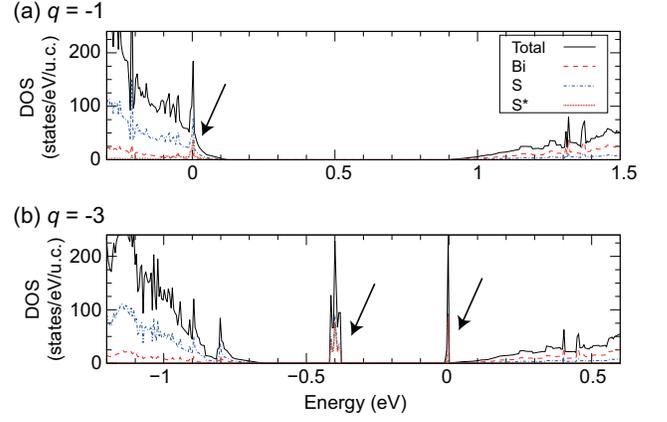}
\caption{Partial DOS for the system with V$_{\mathrm{Bi}}$ with (a) $q=-1$ and (b) $q=-3$. The zero of the energy in the DOS plots is the Fermi energy for each condition. Some of the impurity levels are shown with arrows. S$^*$ shown in the legend denotes S$^{\mathrm{ap}}$ just next to the vacancy site.}
\label{fig:Bidef}
\end{center}
\end{figure}

For the V$_{X}$ defect that shows a change of $q$ in the in-gap energy region, we present the partial DOS for $X=$ Bi with $q=-1, -3$ in Fig.~\ref{fig:Bidef}.
There are very sharp DOS peaks that mainly consist of S$^{\mathrm{ap}}$ just next to the vacancy site. The energy level of this impurity state can lie within the band gap as shown in Fig.~\ref{fig:Bidef}(b), which induces the change of $q$ in Fig.~\ref{fig:cation_both}. To say, $q=-3$ gives electron occupation of such defect states as shown in Fig.~\ref{fig:Bidef}(b).

\subsection{Point defects with $X$/S or S/$X$ substitution\label{sec:XSdefects}}

\begin{figure}
\begin{center}
\includegraphics[width=8.3 cm]{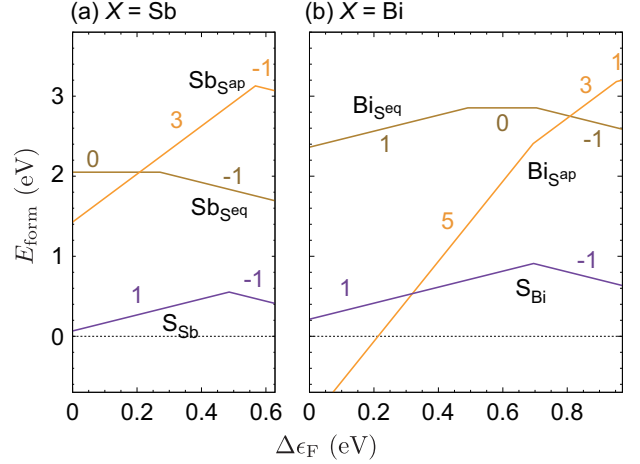}
\caption{Defect formation energies $E_{\mathrm{form}}$ of $X$/S or S/$X$ point defects for (a) LaOSbS$_2$ ($X=$ Sb) and (b) LaOBiS$_2$ ($X=$ Bi). The horizontal line (i.e. the Fermi level) is restricted to the energy range between the band edges corrected by HSE06. The values of $q$, which equals to the slope of each line, are shown beside the line.}
\label{fig:anti_both}
\end{center}
\end{figure}

\begin{figure*}
\begin{center}
\includegraphics[width=16 cm]{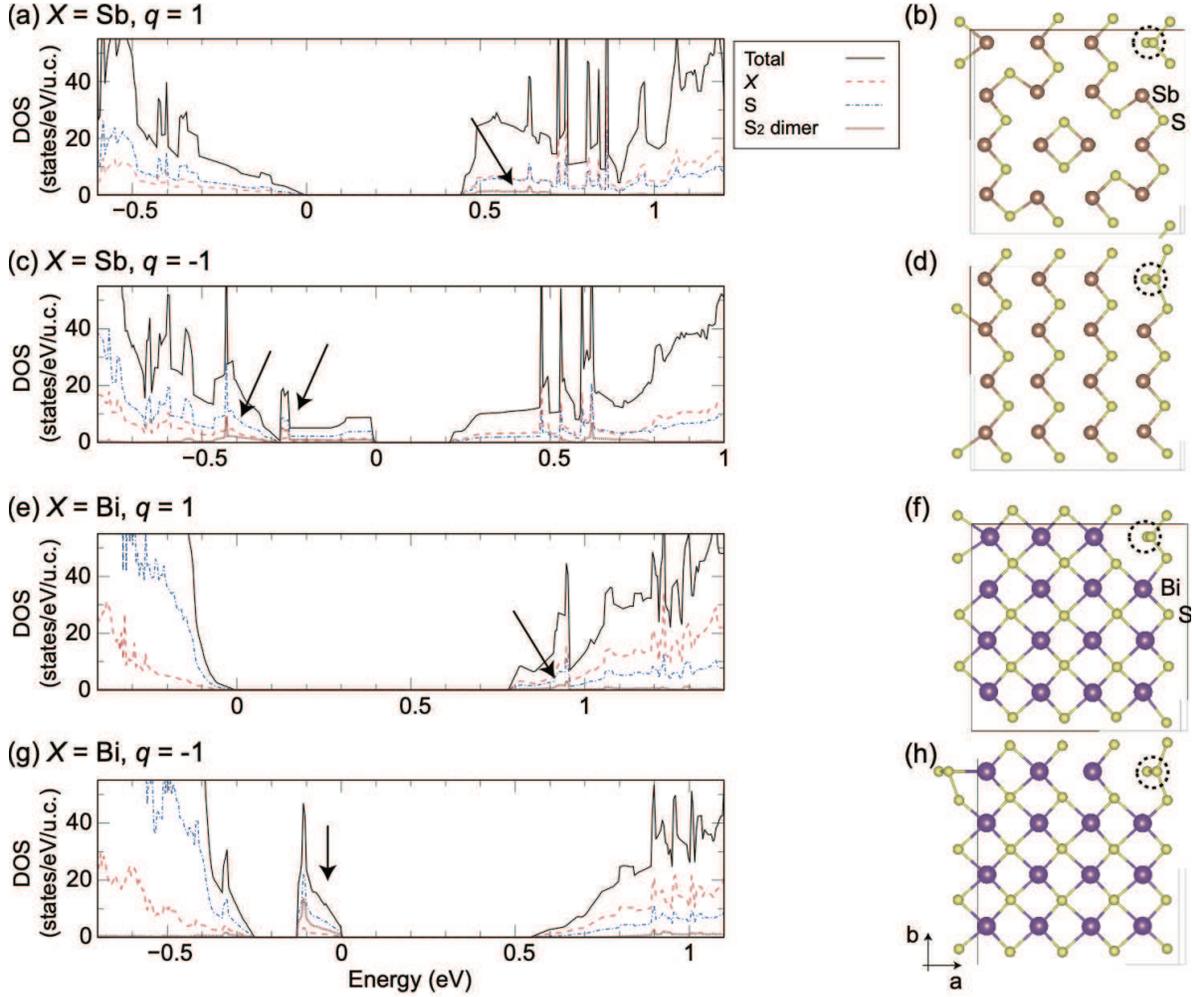}
\caption{(a)(c)(e)(g) Partial DOS and (b)(d)(f)(h) the crystal structure of the $X$S$_2$ plane that includes the S$_{X}$ defect. (a)(b) $X=$ Sb with $q=1$, (c)(d) $X=$ Sb with $q=-1$, (e)(f) $X=$ Bi with $q=1$, and (g)(h) $X=$ Bi with $q=-1$. The zero of the energy in the DOS plots is the highest occupied energy level for each condition. Some of the impurity levels are shown with arrows in the DOS plots. The partial density of states for the S$_2$ dimer surrounded by the black broken lines in the crystal structure are shown in the DOS plots. In the crystal structure, brown, purple, and yellow spheres represent Sb, Bi, and S atoms, respectively.}
\label{fig:DOS_SX}
\end{center}
\end{figure*}

Formation energies of the point defects of $X_{\mathrm{S}}$ or S$_{X}$ are shown in Fig.~\ref{fig:anti_both}.
These defects involve an anion-cation exchange, by which we found that many in-gap states take place.
Therefore, we did not apply the correction $E_{\mathrm{be}}[ D^q ]$ for all the point defects investigated in this section.

We can see that the defect formation energies of $X_{\mathrm{S}}$ are all high at the conduction band bottom, where the chemical potential is considered to lie in these n-type materials.
Thus, we do not go into details about these defects. On the other hand, the defect formation energy of ${\mathrm{S}}_{X}$ is relatively low at the conduction band bottom.
Figure~\ref{fig:DOS_SX} presents the partial DOS and the crystal structure of the $X$S$_2$ plane that includes the S$_{X}$ defect.
In this case, the impurity level shown in Fig.~\ref{fig:DOS_SX}(c)(g) mainly consists of the S$_2$ dimer surrounded by the black broken lines in Fig.~\ref{fig:DOS_SX}(d)(h), respectively.
In Fig.~\ref{fig:DOS_SX}(d)(h), the S$_2$ dimer is more separate from the surrounding atoms than that shown in Fig.~\ref{fig:DOS_SX}(b)(f), as manifested by the bond angle of in-plane sulfurs.
The $q=-1$ charge state allows the occupation of the impurity levels, which seems to stabilize the local structures containing the S$_2$ dimer more separate from the surrounding atoms.
It is also noteworthy that the Sb-S network is again strongly disarranged in Fig.~\ref{fig:DOS_SX}(b), as we have seen in Figs.~\ref{fig:Seqdef}(b)(d).
This also suggests that the SbS$_2$ plane is prone to be disordered.

Since here we considered the $X$-poor and S-rich environment (see Fig.~\ref{fig:chempots}), the ${\mathrm{S}}_{X}$ defect cannot be further stabilized by a different choice of the atomic chemical potentials.
The $X_{\mathrm{S}}$ defects can be stabilized to some extent, but still has positive formation energy even if one considers the $X$-rich and S-poor environment, as shown in Fig.~\ref{fig:anti_sppl}.

\subsection{Fluorine point defects}

\begin{figure}
\begin{center}
\includegraphics[width=8.3 cm]{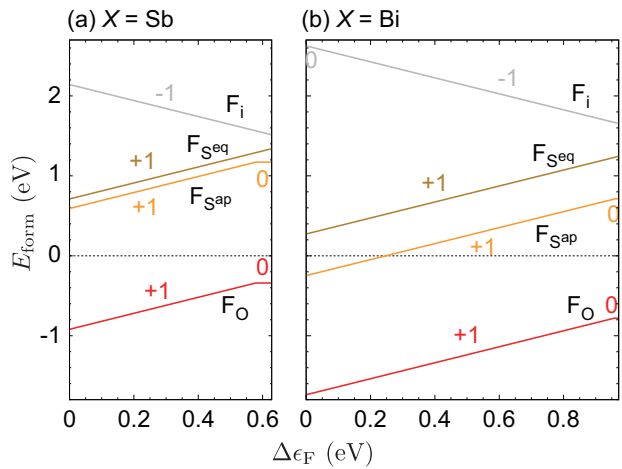}
\caption{Defect formation energies $E_{\mathrm{form}}$ of fluorine point defects for (a) LaOSbS$_2$ ($X=$ Sb) and (b) LaOBiS$_2$ ($X=$ Bi). The horizontal line (i.e. the Fermi level) is restricted to the energy range between the band edges corrected by HSE06. The values of $q$, which equals to the slope of each line, are shown beside the line.}
\label{fig:F_both}
\end{center}
\end{figure}

\begin{figure}
\begin{center}
\includegraphics[width=8.3 cm]{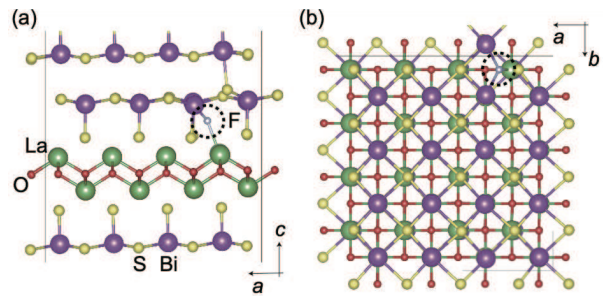}
\caption{The position of the interstitial fluorine atom, F$_{\mathrm{i}}$, is shown in the $4\times 4\times 2$ supercell as (a) side view and (b) top view.
Green, red, purple, yellow, and white spheres represent La, O, Bi, S, and F atoms, respectively. For both panels, atoms far from the fluorine atom are omitted for visibility.}
\label{fig:Finter}
\end{center}
\end{figure}

Finally, to see the effects of fluorine doping, we calculated fluorine-related point defects as shown in Figs.~\ref{fig:F_both}(a)--(b) for LaOSbS$_2$ and LaOBiS$_2$, respectively.
We applied the correction $E_{\mathrm{be}}[ D^q ]$ for all the defects considered here.
For interstitial fluorine doping, F$_{\mathrm{i}}$, Figure~\ref{fig:Finter} shows the most stable position of a fluorine atom we found.
Note that we also found that the defect formation energy of F$_{\mathrm{i}}$ for the case when one places a fluorine atom between two $X$S$_2$ layers is almost similar. The energy difference between that case and the most stable case is at most 0.15 eV for $X=$ Sb and 0.02 eV for $X=$ Bi.

It is clearly presented that, for both compounds, the substitutional doping for oxygen, F$_{\mathrm{O}}$, is the most stable point defect when fluorine is introduced into the crystal.
This situation is unchanged by using different sets of chemical potentials of atoms (see Fig.~\ref{fig:F_sppl}).
Its charge state is basically F$_{\mathrm{O}}$$^{\bullet}$ (i.e., $q=+1$) when the chemical potential lies in the band gap, which is naturally understood by considering that O$^{2-}$ is replaced with F$^{-}$.
This is consistent with many experimental studies of LaOBiS$_2$ where fluorine is doped into the crystal to introduce electron carriers~\cite{BiS2_2}.
The defect formation energy of F$_{\mathrm{O}}$ is higher in $X=$ Sb, which is consistent with the previous theoretical study~\cite{Hirayama}.
Even so, it is noteworthy that the defect formation energy of F$_{\mathrm{O}}$ is negative both for LaOSbS$_2$ and for LaOBiS$_2$.
In fact, an experimental study found that lattice parameters are changed and the electrical conductivity is to some extent increased by fluorine doping for LaOSbSe$_2$~\cite{LnOSbSe2}, which suggests that fluorine was successfully doped into the crystal in experiments.
In the following subsection, we discuss why carrier control is still difficult in Sb compounds, even though the fluorine substitutional doping seems to be realized.

We note that the defect formation energy of F$_{\mathrm{O}}$ is close to zero at the conduction band bottom, in particular for $X=$ Sb, which can be positive by using different chemical potentials of atoms, as shown in Fig.~\ref{fig:F_sppl}(b). In addition, the calculation results, of course, should be different for different compounds (i.e., different lanthanoid and chalcogen elements).
In this sense, our calculation results suggest that fluorine substitutional doping for oxygen can be energetically unfavorable in some $X=$ Sb compounds or some experimental setup.
  
\subsection{Differences between $X=$ Sb and Bi compounds\label{sec:Sbcarrier}}

Here, we summarize the important differences between $X=$ Sb and Bi compounds:
(i) V$_{\mathrm{O}}$ has a lower formation energy in $X=$ Bi than in $X=$ Sb.
(ii) In-plane structural instability seems to result in the relatively stable V$_{\mathrm{S}^{\mathrm{eq}}}$ for $X=$ Sb (and S$_{X}$).
The in-plane $X$-S bonding network is more strongly disarranged for $X=$ Sb.
(iii) F$_{\mathrm{O}}$ has a lower formation energy in $X=$ Bi than in $X=$ Sb.

Both (i) and (iii) suggest that the electron carriers are much more difficult to introduce because of higher formation energies of corresponding point defects in $X=$ Sb compounds than in $X=$ Bi compounds.
Even when one succeeds in introducing the electron carriers, the strong in-plane structural instability in $X=$ Sb can be an obstacle for realizing high electrical conductivity, as suggested by (ii).
While we have only investigated the point defects in this study, several possible higher-order defects, such as a twin defect, can occur in these compounds.
It is an important future problem to clarify the origins of the relatively low electrical conductivity in $X=$ Sb compounds from this viewpoint.

\subsection{Some remarks for effective carrier control}
Our calculation suggests that the anion vacancy (V$_{\mathrm{O}}$ or V$_{\mathrm{S}}$) can be relatively stable in some chemical environment. This observation means that oxygen-rich and sulfur-rich environment can be beneficial to synthesize the crystal with a low concentration of defects. It also means that one can possibly control the electron carrier concentration by changing the chemical potentials of oxygen and sulfur. When one would like to introduce a large amount of electron carriers, the fluorine substitutional doping F$_{\mathrm{O}}$ is shown to be effective as is well known in experimental studies of BiS$_2$ compounds.

\section{Conclusion\label{sec:sum}}

In this paper, we have systematically investigated the defect formation energy of several point defects in LaO$X$S$_2$ ($X=$ Sb, Bi) using first-principles calculation.
We have found that anion replacements S$_\mathrm{O}$ and O$_\mathrm{S}$ are not stable while V$_{\mathrm{O}}$ and V$_{\mathrm{S}}$ can take place, while the formation energy of V$_{\mathrm{O}}$ is higher in $X=$ Sb than in $X=$ Bi.
It is characteristic that V$_{\mathrm{S}^{\mathrm{eq}}}$ becomes much more stable in $X=$ Sb than in $X=$ Bi, due to the formation of an Sb$_2$ dimer and the occupation of the impurity energy levels.
The formation energies of cation defects, $X_{\mathrm{S}}$, and S$_{X}$ are positive for the atomic chemical potentials used in this study.
Fluorine likely replaces oxygen for both $X=$ Sb and Bi. The defect formation energy of F$_{\mathrm{O}}$ is negative for both compounds, while that for $X=$ Sb is much higher than $X=$ Bi.
Our study has clarified the stability of several point defects and suggested that the in-plane structural instability is enhanced in $X=$ Sb.
This knowledge should be helpful for understanding and controlling the transport properties of LaO$X$S$_2$ and related compounds by impurity doping.

\acknowledgments
We appreciate the fruitful discussion with Yosuke Goto, Yoshikazu Mizuguchi, and Kazutaka Nishiguchi.
This study was supported by JST CREST (No.~JPMJCR20Q4), Japan.
Part of the numerical calculations were performed using the large-scale computer systems provided by the supercomputer center of the Institute for Solid State Physics, the University of Tokyo, and the Information Technology Center, the University of Tokyo.

\section*{Appendix A: Density of states for perfect crystal}

\begin{figure}
\begin{center}
\includegraphics[width=8.3 cm]{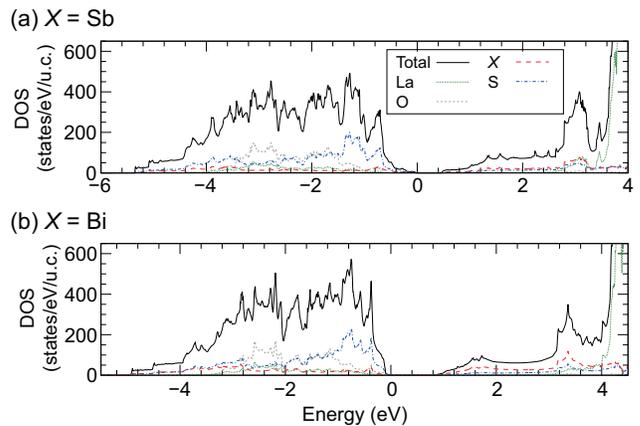}
\caption{Partial DOS for perfect crystal of LaO$X$S$_2$: (a) $X=$ Sb and (b) $X=$ Bi.
To compare them with those shown in the main text, DOS is normalized as that for the $4\times 4\times 2$ supercell.
To say, u.c. = unit cell used for the $4\times 4\times 2$ supercell calculation containing 320 atoms.}
\label{fig:DOS_nondoped}
\end{center}
\end{figure}

Figure~\ref{fig:DOS_nondoped} presents the partial DOS for perfect crystal of LaO$X$S$_2$ ($X=$ Sb, Bi).

\section*{Appendix B: Extended data for calculating chemical potentials of atoms}

Table~\ref{tab:list_mater} presents a list of solid compounds, which in addition to LaOSbS$_2$, were used to determine the chemical potentials of atoms.
Since we performed the first-principles calculation at zero temperature, we basically adopted the space group of the low-temperature phase for each compound.
For LaOF~\cite{LaOF}, we compared the total energies of $\beta$- (rhombohedral, $R\bar{3}m$) and t- (tetragonal, $P4/nmm$ phases, and found that the latter has a bit lower total energy. Thus, we adopted the tetragonal phase here.

Using Table~\ref{tab:list_mater}, we obtained the chemical potentials of atoms as listed on Tables~\ref{tab:equil_Sb} and \ref{tab:equil_Bi}, for LaOSbS$_2$ and LaOBiS$_2$, respectively. 
As described in Sec.~\ref{sec:chem}, we defined $\Delta \mu_i \equiv \mu_i - \mu_i^0$ for each atom $i$, where $\mu_i^0$ is the total energy per atom for La (solid), O$_2$ molecule, F$_2$ molecule, Sb (solid), Bi (solid), and $\alpha$-S (solid), for La, O, F, Sb, Bi, and S, respectively.
For LaOBiS$_2$, a set of solid compounds with $\Delta_{\mathrm{LaOBiS}_2}\equiv E [\mathrm{LaOBiS}_2] - (\mu [\mathrm{La}] + \mu [\mathrm{O}] + \mu [\mathrm{Bi}] + 2\mu [\mathrm{S}] ) > 100$ meV were discarded, because the chemical equilibrium where LaOBiS$_2$ is much unstable is not appropriate for our purpose.
The A1 on Table~\ref{tab:equil_Sb} and the B1 on Table~\ref{tab:equil_Bi} were adopted in our theoretical analysis shown in the main text.

\begin{table}
\caption{A list of solid compounds used to determine the chemical potentials of atoms.}
\label{tab:list_mater}
\begin{tabular}{c c c}
\hline
 Compound & Space group & Reference\\
 \hline \hline
 La & $P6_3/mmc$ & [\onlinecite{Bi_text}]\\
S & $Fddd$ & [\onlinecite{alphaS}] \\
La$_2$O$_3$ & $P\bar{3}m$1 & [\onlinecite{La2O3}] \\
LaF$_3$ & $P\bar{3}c$1 & [\onlinecite{LaF3}] \\
La$_2$S$_3$ & $Pnma$ & [\onlinecite{La2S3}] \\
LaOF & $P4/nmm$ & [\onlinecite{LaOF}]\\
La$_2$O$_2$S$_2$ & $Cmca$ & [\onlinecite{La2O2S2}] \\
La$_2$O$_2$S &  $P\bar{3}m$1 & [\onlinecite{La2O2S}]\\
La$_2$O$_2$SO$_4$ & $C2/c$ & [\onlinecite{La2O2SO4}]\\
\hline
Bi & $R\bar{3}m$ & [\onlinecite{Bi_text}] \\ 
LaBi & $Fm\bar{3}m$ & [\onlinecite{LaBi}] \\
La$_2$Bi & $I4/mmm$ &  [\onlinecite{LaBi}] \\
La$_4$Bi$_3$ & $I\bar{4}3d$ & [\onlinecite{La4Pn3}] \\
La$_5$Bi$_3$ & $P63/mcm$ &  [\onlinecite{LaBi}] \\
Bi$_2$O$_3$ & $P2_1/c$ & [\onlinecite{Bi2O3}]\\
Bi$_2$O$_4$ & $C$2/$c$ & [\onlinecite{Bi2O4}]\\
Bi$_4$O$_7$ & $P\bar{1}$ & [\onlinecite{Bi4O7}] \\
BiF$_3$ & $Fm\bar{3}m$ & [\onlinecite{BiF3}]\\
Bi$_2$S$_3$ & $Pnma$ & [\onlinecite{Bi2S3}]\\
BiOF & $P$4/$nmm$ & [\onlinecite{BiOF}]\\
Bi$_2$OS$_2$ & $P4/nmm$ & [\onlinecite{LaOBiS2_gap}]\\
Bi$_2$O$_2$S & $Pnnm$ & [\onlinecite{Bi2O2S}] \\
Bi$_2$(SO$_4$)$_3$ & $C2/c$ & [\onlinecite{Bi2SO4_3}]\\
\hline
Sb & $R\bar{3}m$ & [\onlinecite{Bi_text}] \\
LaSb & $Fm\bar{3}m$ & [\onlinecite{LaSb}]\\
LaSb$_2$ & $Cmca$ & [\onlinecite{LaSb2}] \\
La$_2$Sb & $I4/mmm$ & [\onlinecite{La2Sb}] \\
La$_4$Sb$_3$ & $I\bar{4}3d$ & [\onlinecite{La4Pn3}] \\
La$_5$Sb$_3$ & $P63/mcm$ & [\onlinecite{La5Sb3}] \\
Sb$_2$O$_3$ & $Fd\bar{3}m$ & [\onlinecite{Sb2O3}] \\
Sb$_2$O$_4$ & $Pna2_1$ & [\onlinecite{Sb2O4}] \\
Sb$_2$O$_5$ & $C$2/$c$ & [\onlinecite{Sb2O5}] \\
SbF$_3$ & $Ama2$ &[\onlinecite{SbF3}] \\
Sb$_2$S$_3$ & $Pnma$ & [\onlinecite{Bi2S3}]\\
SbOF & $Pnma$ & [\onlinecite{SbOF}] \\
Sb$_3$O$_4$F & $P2_1/c$ & [\onlinecite{Sboxyflu}]\\
Sb$_3$O$_2$F$_5$ & $P2/c$ & [\onlinecite{Sboxyflu}]\\
Sb$_2$OS$_2$ & $P\bar{1}$ & [\onlinecite{Sb2OS2}] \\
Sb$_2$(SO$_4$)$_3$ & $P2_1/c$ & [\onlinecite{Sb2SO4_3}]\\
Sb$_2$O(SO$_4$)$_2$ & $P4_1 2_1 2$ & [\onlinecite{Sb2O_SO4_2}] \\
 \hline
 \end{tabular}
 \end{table}

\begin{table*}
\caption{A list of chemical potentials of atoms determined by chemical equilibrium conditions as described in the main text for $X=$ Sb. Chemical potentials are shown in eV.}
\label{tab:equil_Sb}
\scalebox{0.92}{
\begin{tabular}{c c c c c c c c c c c c c c c c c c}
\hline \hline
Subset & $\Delta\mu [\mathrm{La}]$ & $\Delta\mu [\mathrm{O}]$ & $\Delta\mu [\mathrm{F}]$ & $\Delta\mu[\mathrm{Sb}]$ & $\Delta\mu [\mathrm{S}]$ & \multicolumn{12}{c}{Coexisting compounds}\\
 & & & & & & LaF$_3$ & LaOF & La$_2$S$_3$ & La$_2$O$_2$S$_2$ & La$_2$O$_2$S & La$_2$O$_2$SO$_4$ & Sb$_2$S$_3$ & Sb$_2$OS$_2$ & Sb$_2$O$_3$ & Sb & S & LaOSbS$_2$\\
 \hline
 A1 & $-5.49$ & $-2.50$ & $-3.76$ & $-0.68$ & $-0.01$ & $\checkmark$ & & $\checkmark$ & $\checkmark$ & $\checkmark$ & & & & & &  & $\checkmark$\\
 A2 & $-5.50$ & $-2.49$ & $-3.75$ & $-0.69$ &  $0.00$ & $\checkmark$ & & $\checkmark$ & $\checkmark$ & & & & & & & $\checkmark$ & $\checkmark$\\
 A3 & $-4.81$ & $-2.95$ & $-3.98$ &  $0.00$ & $-0.46$ & $\checkmark$ & & $\checkmark$ & & $\checkmark$ & & & & & $\checkmark$ & & $\checkmark$\\
 A4 & $-5.50$ & $-2.54$ & $-3.75$ & $-0.64$ &  $0.00$ & $\checkmark$ & & $\checkmark$ & & & & $\checkmark$ & & & & $\checkmark$ & $\checkmark$\\ 
 A5& $-4.86$ & $-2.97$ & $-3.96$ &  $0.00$ & $-0.43$ & $\checkmark$ &  & $\checkmark$ & & & & $\checkmark$ & & & $\checkmark$ &  & $\checkmark$\\
 A6 & $-5.53$ & $-2.23$ & $-3.95$ &  $0.00$ & $-0.46$ & & $\checkmark$ & &  & $\checkmark$ & & & & $\checkmark$ & $\checkmark$ &  & $\checkmark$\\
 A7 & $-5.60$ & $-2.23$ & $-3.87$ &  $0.00$ & $-0.43$ & & $\checkmark$ & & & & & $\checkmark$ & $\checkmark$ & & $\checkmark$ &  & $\checkmark$\\
 A8 & $-5.60$ & $-2.23$ & $-3.88$ &  $0.00$ & $-0.43$ & & $\checkmark$ & & & & & & $\checkmark$ & $\checkmark$ & $\checkmark$ &  & $\checkmark$\\
 A9 & $-5.79$ & $-2.20$ & $-3.72$ & $-0.68$ & $-0.01$ & & $\checkmark$ & & $\checkmark$ & $\checkmark$ & $\checkmark$ & & & & &  & $\checkmark$\\
A10 & $-5.80$ & $-2.20$ & $-3.71$ & $-0.69$ & $0.00$ & & $\checkmark$ & & $\checkmark$ & & $\checkmark$ & & & & & $\checkmark$ & $\checkmark$\\
A11 & $-5.57$ & $-2.20$ & $-3.93$ & $-0.04$ & $-0.44$ & & $\checkmark$ & & & $\checkmark$ & $\checkmark$ & & & $\checkmark$ & & & $\checkmark$\\
A12 & $-5.87$ & $-2.17$ & $-3.66$ & $-0.64$ &  $0.00$ & & $\checkmark$ & & & & $\checkmark$ & $\checkmark$ & & & & $\checkmark$ & $\checkmark$\\
A13 & $-5.69$ & $-2.17$ & $-3.84$ & $-0.09$ & $-0.37$ & & $\checkmark$ & & & & $\checkmark$ & $\checkmark$  &$\checkmark$ &  & &  & $\checkmark$\\
A14 & $-5.68$ & $-2.18$ & $-3.85$ & $-0.08$ & $-0.38$ & & $\checkmark$ & & & & $\checkmark$ & & $\checkmark$ & $\checkmark$ & &  & $\checkmark$\\
\hline \hline
 \end{tabular}
 }
 \end{table*}
 
 \begin{table*}
\caption{A list of chemical potentials of atoms determined by chemical equilibrium conditions as described in the main text for $X=$ Bi. For each set of compounds, $\Delta_{\mathrm{LaOBiS}_2}\equiv E [\mathrm{LaOBiS}_2] - (\mu [\mathrm{La}] + \mu [\mathrm{O}] + \mu [\mathrm{Bi}] + 2\mu [\mathrm{S}] )$ is also shown. All values are shown in eV.}
\label{tab:equil_Bi}
\begin{tabular}{c c c c c c c c c c c c c c c c c c}
\hline \hline
Subset  & $\Delta\mu [\mathrm{La}]$ & $\Delta\mu [\mathrm{O}]$ & $\Delta\mu [\mathrm{F}]$ & $\Delta\mu[\mathrm{Bi}]$ & $\Delta\mu [\mathrm{S}]$ & $\Delta_{\mathrm{LaOBiS}_2}$ & \multicolumn{10}{c}{Coexisting compounds}\\
 & & & & & & & LaF$_3$ & LaOF & La$_2$S$_3$ & La$_2$O$_2$S$_2$ & La$_2$O$_2$S & La$_2$O$_2$SO$_4$ & Bi$_2$S$_3$ & Bi$_2$OS$_2$ & Bi & S\\
 \hline
B1 & $-5.49$ & $-2.50$ & $-3.76$ & $-0.86$ & $-0.01$ & $0.02$ & $\checkmark$ & & $\checkmark$ & $\checkmark$  & $\checkmark$  & & $\checkmark$ &  &  & \\
B2 & $-4.63$ & $-3.07$ & $-4.04$ & $0.00$ & $-0.59$ & $0.02$ & $\checkmark$ & & $\checkmark$ & & $\checkmark$ & & $\checkmark$  &  & $\checkmark$ & \\
B3 & $-5.50$ & $-2.49$ & $-3.75$ & $-0.88$ & $0.00$ & $0.03$ & $\checkmark$ & & $\checkmark$ & $\checkmark$  & & & $\checkmark$  &  &  & $\checkmark$  \\
B4 & $-5.50$ & $-2.20$ & $-4.01$ & $0.00$ & $-0.59$ & $0.02$ & & $\checkmark$ & & & $\checkmark$ & $\checkmark$ & $\checkmark$ &  & $\checkmark$ & \\
B5 & $-5.79$ & $-2.20$ & $-3.72$ & $-0.86$ & $-0.01$ & $0.02$ & & $\checkmark$ & & $\checkmark$ &$\checkmark$ & $\checkmark$ & $\checkmark$ &  &  & \\
B6 & $-5.50$ & $-2.20$ & $-4.00$ & $0.00$ & $-0.59$ & $0.02$ & & $\checkmark$& & &  &$\checkmark$ & $\checkmark$  & $\checkmark$  & $\checkmark$ & \\
B7 & $-5.80$ & $-2.20$ & $-3.71$ & $-0.88$ & $0.00$ & $0.03$ & & $\checkmark$ &  & $\checkmark$  & & $\checkmark$ &$\checkmark$  &  &  & $\checkmark$ \\
\hline \hline
 \end{tabular}
 \end{table*}

\section*{Appendix C: Defect formation energy calculated with different sets of chemical potentials}
 
\begin{figure*}
\begin{center}
\includegraphics[width=16.5 cm]{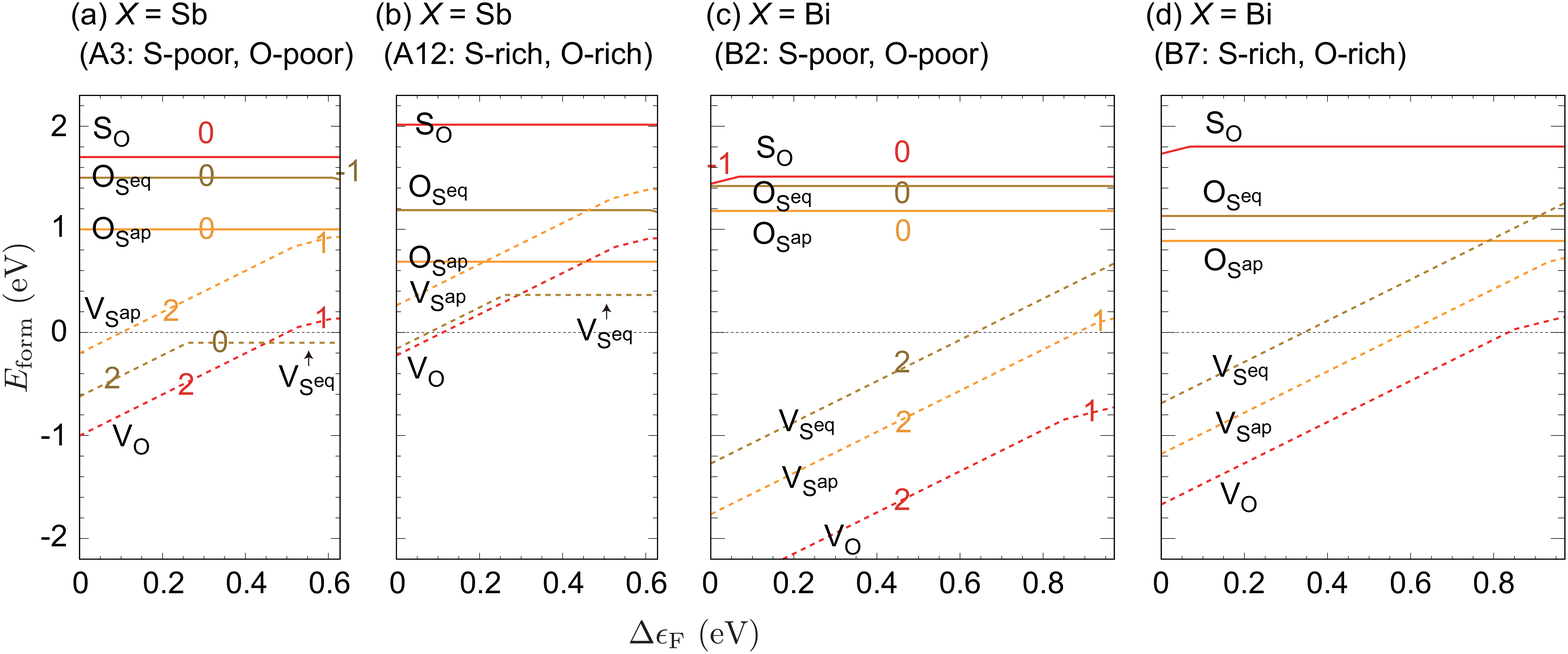}
\caption{Defect formation energies $E_{\mathrm{form}}$ of anion point defects for (a)(b) LaOSbS$_2$ ($X=$ Sb) and (c)(d) LaOBiS$_2$ ($X=$ Bi). The horizontal line is restricted to the energy range between the band edges corrected by HSE06. The values of $q$, which equals to the slope of each line, are shown beside the line. Sets of the chemical potentials of atoms, A3, A12 shown in Table~\ref{tab:equil_Sb} and B2, B7 shown in Table~\ref{tab:equil_Bi} are used for panels (a)--(d), respectively.}
\label{fig:anion_sppl}
\end{center}
\end{figure*}

\begin{figure*}
\begin{center}
\includegraphics[width=16.5 cm]{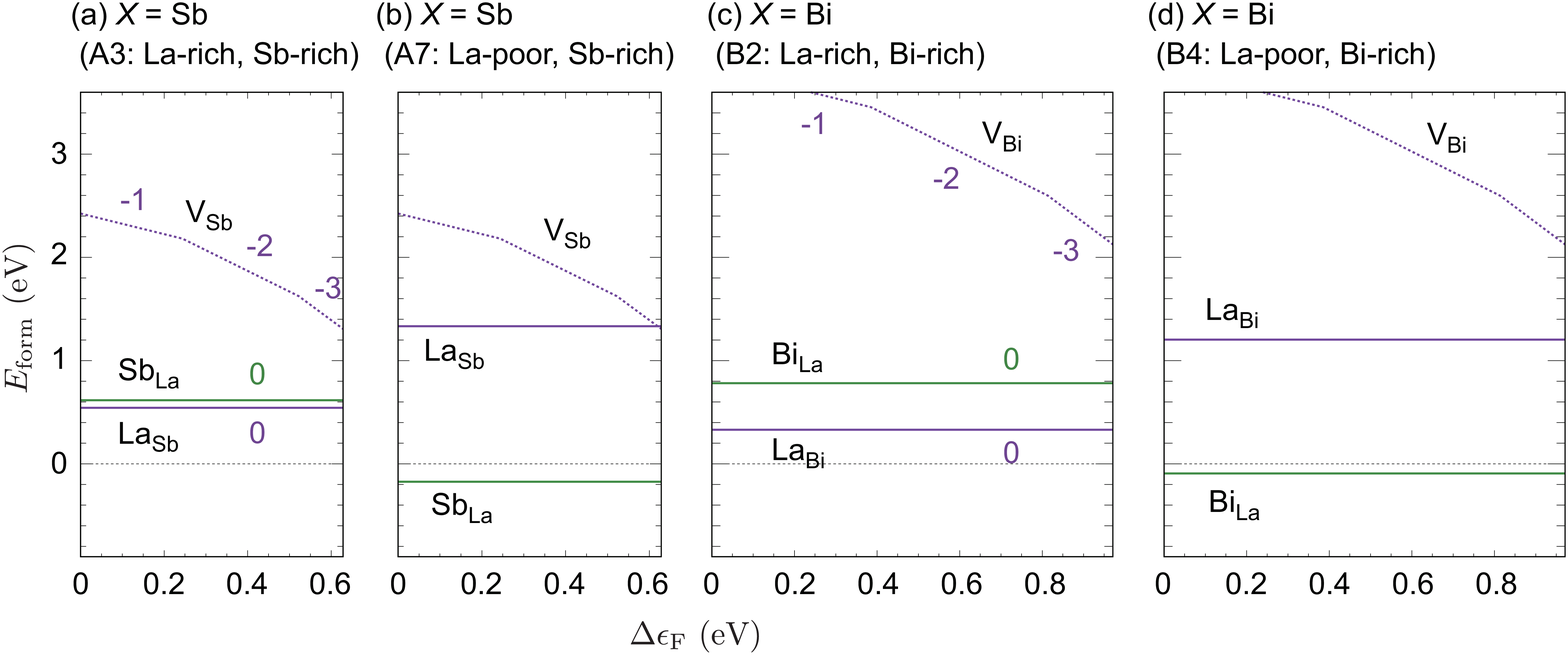}
\caption{Defect formation energies $E_{\mathrm{form}}$ of cation point defects for (a)(b) LaOSbS$_2$ ($X=$ Sb) and (c)(d) LaOBiS$_2$ ($X=$ Bi). The horizontal line is restricted to the energy range between the band edges corrected by HSE06. The values of $q$, which equals to the slope of each line, are shown beside the line. Sets of the chemical potentials of atoms, A3, A7 shown in Table~\ref{tab:equil_Sb} and B2, B4 shown in Table~\ref{tab:equil_Bi} are used for panels (a)--(d), respectively.}
\label{fig:cation_sppl}
\end{center}
\end{figure*}

\begin{figure}
\begin{center}
\includegraphics[width=8.4 cm]{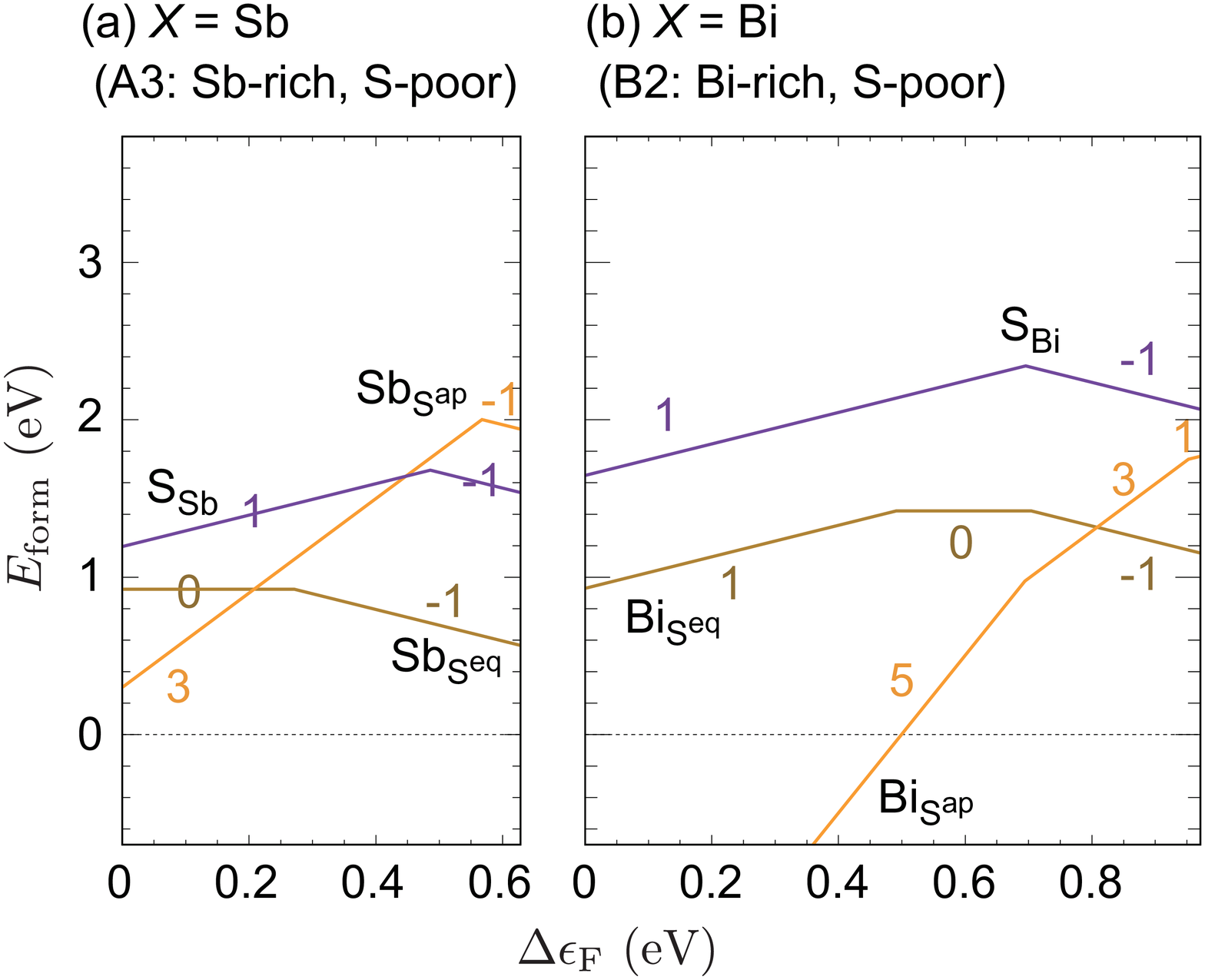}
\caption{Defect formation energies $E_{\mathrm{form}}$ of $X$/S or S/$X$ point defects for (a) LaOSbS$_2$ ($X=$ Sb) and (b) LaOBiS$_2$ ($X=$ Bi). The horizontal line is restricted to the energy range between the band edges corrected by HSE06. The values of $q$, which equals to the slope of each line, are shown beside the line. Sets of the chemical potentials of atoms, A3 shown in Table~\ref{tab:equil_Sb} and B2 shown in Table~\ref{tab:equil_Bi} are used for panels (a)--(d), respectively.}
\label{fig:anti_sppl}
\end{center}
\end{figure}

\begin{figure*}
\begin{center}
\includegraphics[width=16.5 cm]{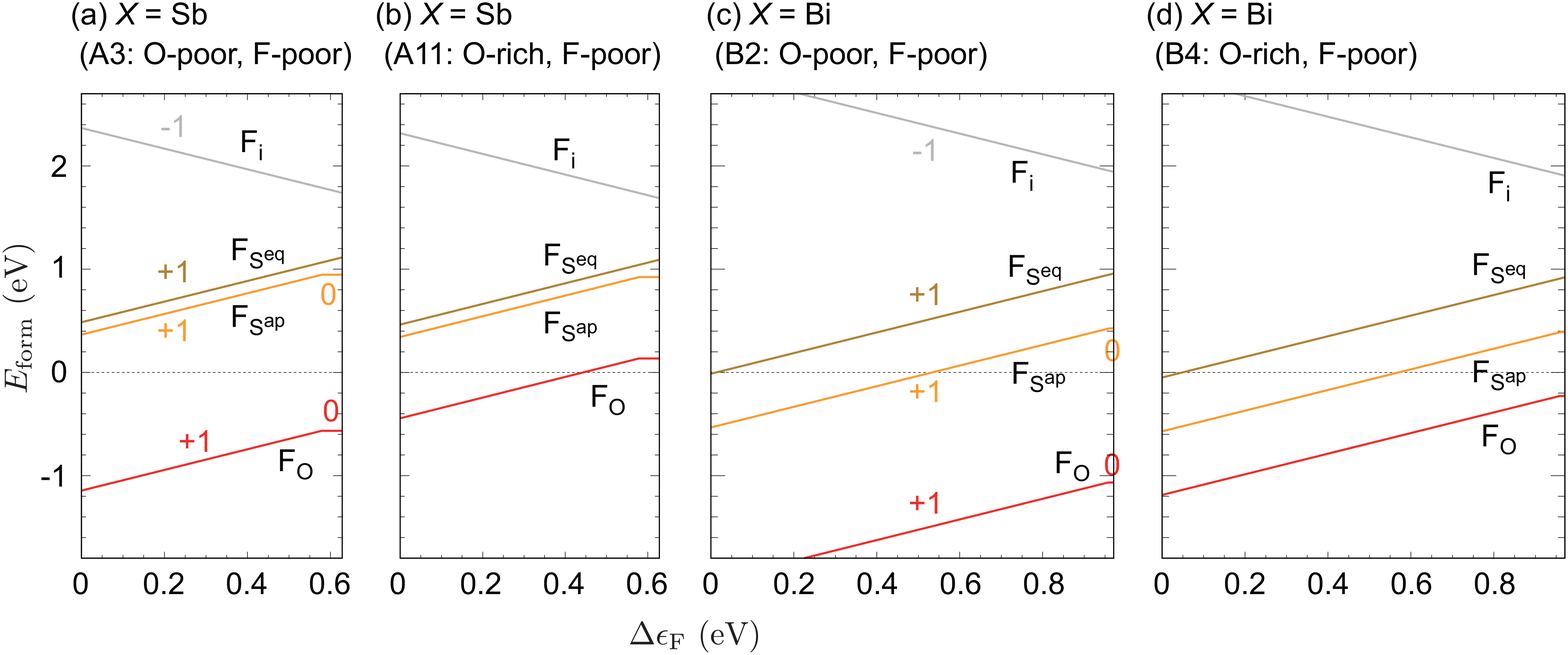}
\caption{Defect formation energies $E_{\mathrm{form}}$ of fluorine point defects for (a)(b) LaOSbS$_2$ ($X=$ Sb) and (c)(d) LaOBiS$_2$ ($X=$ Bi). The horizontal line is restricted to the energy range between the band edges corrected by HSE06. The values of $q$, which equals to the slope of each line, are shown beside the line. Sets of the chemical potentials of atoms, A3, A11 shown in Table~\ref{tab:equil_Sb} and B2, B4 shown in Table~\ref{tab:equil_Bi} are used for panels (a)--(d), respectively.}
\label{fig:F_sppl}
\end{center}
\end{figure*}

Defect formation energy calculated with different sets of chemical potentials are shown in Figs.~\ref{fig:anion_sppl}--\ref{fig:F_sppl}.

\end{document}